\documentclass[aps,prl,showpacs,twocolumn]{revtex4}
\usepackage{latexsym}
\usepackage{amssymb}
\usepackage{graphicx}
\begin{document}
\title{Vanishing of the Dissipationless Spin Hall Effect in
a Diffusive Two-Dimensional Electron Gas with Spin-Orbit Coupling}
\author{S. Y. Liu}
\author{X. L. Lei}
\affiliation{Department of Physics, Shanghai Jilting University,
1954 Huashan Road, Shanghai 200030, China}
\date{\today}
\begin{abstract}
We propose a nonequilibrium Green's function approach to study the
spin-Hall effect in a two-dimensional electron system with both
the Rashba and Dresselhaus spin-orbit couplings. By taking into
account the long-range electron-impurity scattering, the derived
kinetic equations are solved numerically. It is found the
vanishing of the total zero-temperature dissipationless spin-Hall
effect, contributing from the intrinsic and disorder-mediated
processes. This result has been examined in the wide ranges of
spin-orbit coupling constants and electron density.
\end{abstract}

\maketitle

In the growing field of spintronics, which aims to the
manipulation of spin degrees of freedom in
semiconductors\cite{Sarma}, the issue of creating spin
polarization of carriers in nonmagnetic semiconductors with
spin-orbit coupling by employing only the electric field has
attracted a great deal of theoretical and experimental attention.
More recently, the non-dissipative spin currents with the
spin-polarization perpendicular to the flow direction and to the
applied electric field have been predicted in p-doped bulk
semiconductor\cite{Zhang} and two-dimensional (2D) systems with
the Rashba \cite{Sinova} and Dresselhaus spin-orbit
couplings\cite{Shen, Sinitsyn}. This dissipationless spin-Hall
effect essentially arises from the dc-field-induced transition
between the helicity bands: the dc field causes the variation of
the electron momentum, but leaves its spin unchanged\cite{Liu}. In
experiment, the spin-Hall effects have been observed in n-doped
bulk semiconductors\cite{Kato} and two-dimensional hole
gas\cite{Wunderlich}.

The disorder effects on the spin-Hall current have been
extensively investigated for the two-dimensional electron systems
with Rashba spin-orbit coupling. In Ref.\,\cite{Schliemann,
Burkov, Nomura}, it has been argued that the disorder can reduce
the spin-Hall effect and the ballistic value of spin-Hall
conductivity reaches for the weak scattering limit. However, based
on the Kubo formalism, Inoue {\it et al.}\cite{Inoue},
Dimitrova\cite{Dimitrova} and Chalaev and Loss\cite{Chalaev} have
shown that the spin-Hall effect disappears even when the disorder
is weak. This conclusion has been confirmed in
Refs.\,\cite{Mischenko} and \cite{Khaetskii} by employing the
spin-density matrix method and the Keldysh approach, respectively.
Further, using the nonequilibrium Green's function approach, Liu
and Lei have demonstrated that, even for the case of long-range
disorders, the dissipationless spin-Hall effect also
vanishes\cite{Liu}. Obviously, the nonexistence of dissipationless
spin-Hall effect still holds for the Dresselhaus spin-orbit
coupling, since it relates to the Rashba coupling through a global
unitary transformation\cite{Shen}. It is commonly believed that
this complete cancellation of spin-Hall effect is not a
consequence of any symmetries\cite{Murakami} and relates to the
isotropy of the dispersion in the helicity basis.

It is very interesting to examine the spin-Hall effect for 2D
electron systems with both the Rashba and Dresselhaus spin-orbit
interactions. In such systems, the spin-orbit-coupled bands become
anisotropic, leading to many interesting phenomena, such as the
spin splitting\cite{SSp}, spin precession and
relaxation\cite{SPR,SPR1,Schliemann1,SPR2}, and anisotropic
transport\cite{Schliemann2}, {\it etc}. In particular, when the
Rashba coefficient is turned to be equal to the Dresselhaus one,
such system can be used as the new type of spin field-effect
transistor (SFET) operated in the nonballistic
regime\cite{Schliemann1}. However, the spin-Hall effect in the 2D
system with the Rashba and Dresselhaus spin-orbit couplings has
been studied only in the ballistic regime\cite{Shen, Sinitsyn}. It
has been demonstrated that the spin-Hall conductivity has a
universal value, independent on the strength of the coupling, but
its sign is determined by the relative ratio of the two couplings.

In this letter, we develop a nonequilibrium Green's function
approach to the spin-Hall effect in 2D electron systems with both
the Rashba and Dresselhaus spin-orbit couplings. The
electron-impurity scattering is considered in the self-consistent
Born approximation and the obtained equations are solved
numerically. We find that the dissipationless spin-Hall
conductivity vanishes at zero temperature. This result contrasts
with the general belief that the anisotropy can lead to the
nonzero dissipationless spin-Hall conductivity in 2D electron
system with spin-orbit coupling\cite{Inoue}.

We consider an effective Hamiltonian for a two-dimensional
electron of momentum ${\bf p}\equiv (p_x,p_y)$ and effective mass
$m$
\begin{equation}
{\tilde H}_0=\frac{{\bf p}^2}{2m}+{\tilde H}_R+{\tilde H}_D,\label{Ham}
\end{equation}
with the terms of Rashba spin-orbit coupling ${\tilde H}_R$ \cite{Rashba},
\begin{equation}
{\tilde H}_R=\alpha(p_y\sigma_x-p_x\sigma_y),
\end{equation}
and Dresselhaus spin-orbit interaction\cite{Dresselhaus},
\begin{equation}
{\tilde H}_D=\beta(p_y\sigma_y-p_x\sigma_x).
\end{equation}
Here, ${\bf \sigma}\equiv (\sigma_x,\sigma_y,\sigma_z)$ are the Pauli matrices,
and, the $\alpha$ and $\beta$ are the coupling constants.  After taking the local
unitary transformation
\begin{equation}
U(\bf p)=\frac 1 {\sqrt{2}}\left (
\begin{array}{cc}
1&1\\
-{\rm e}^{i\chi_{\bf p}}&{\rm e}^{i\chi_{\bf p}}
\end{array}
\right )\label{Uni}
\end{equation}
with the $\chi_{\bf p}$ being
\begin{equation}
\chi_{\bf p}={\rm arg}[\alpha p_y-\beta p_x-i(\alpha p_x-\beta p_y)],
\end{equation}
we can diagonalize this Hamiltonian in the helicity basis $H={\rm
diag}(\varepsilon_{1}(\bf p), \varepsilon_{2}({\bf p}))$. The
$\varepsilon_\mu ({\bf p})$ ($\mu=1,2$) are of the forms
\begin{equation}
\varepsilon_{\mu}({\bf p})=\frac{{\bf p}^2}{2m}+(-1)^\mu \varepsilon_{RD},
\end{equation}
where, $\varepsilon_{RD}\equiv\sqrt{(\alpha p_y-\beta p_x)^2+
(\alpha p_x-\beta p_y)^2}$.

When the system is driven by a dc field ${\bf E}$ applied along
the $x$-direction, the $i$-direction-polarized spin current for
electron with momentum ${\bf p}$ can be defined as ${\bf
j}^i=[\sigma_i {\bf v}+{\bf v} \sigma_i]/4$ with the electron
velocity ${\bf v}=-i[{\bf r},H]$ \cite{Rashba2}. In the helicity
basis, the total $z$-direction-spin current along the $y$ axis,
corresponding to the spin-Hall current,
can be written as
\begin{equation}
J_y^z=\sum_{\bf p} \frac{p_y}{2m}[\rho_{12}({\bf
p})+\rho_{21}({\bf p})] =\sum_{\bf p} \frac{p_y}{m}{\rm Re}
\rho_{12}({\bf p}),\label{Jyz}
\end{equation}
where $\rho_{\mu\nu}({\bf p})$ is the distribution function with
the symmetry relation $\rho_{12}({\bf p})=\rho_{21}^*({\bf
p})$\cite{Jauho}. Further, the spin-Hall conductivity is given by
$\sigma_{sH}=J_y^z/E$.

To derive the kinetic equations for the 2D systems with the
spin-orbit coupling, we follow the procedure described in
Ref.\,\cite{Liu}. We first carry out the Dyson equations for the
real-space less Green's functions in the spin basis. While the
Fourier and the unitary transformation (\ref{Uni}) are performed,
these equations reduce to the kinetic equations in the helicity
basis,
\begin{widetext}
\begin{equation}
\left \{i\frac{\partial}{\partial T}+ie{\bf E}\cdot \nabla_{\bf
p}\right \}{\rm G}^< +\frac {e {\bf E}}{2}\cdot \nabla_{\bf p}
\chi_{\bf p} [{\rm G}^<,\sigma_x]-\varepsilon_{RD}[{\rm
G}^<,\sigma_z]={\Sigma}^r {\rm G}^<+{\Sigma}^< {\rm G}^a -{\rm
G}^r {\Sigma}^<-{\rm G}^< {\Sigma}^a,
\end{equation}
here, the self-energy for the electron-impurity interaction with
scattering matrix $V({\bf p}-{\bf k})$ reads
\begin{equation}
\Sigma^{r,<}({\bf p},T,t)=\frac 12 n_i\sum_{{\bf
k}}|V({\bf p}-{\bf k})|^2\left \{
a_1 {\rm G}^{r,<}+a_2\sigma_x{\rm G}^{r,<}\sigma_x-ia_3[\sigma_x,{\rm G}^{r,<}]\right \},
\end{equation}
\end{widetext}
with the center-of-mass and difference times $T$ and $t$ relating
to the two times of the Green's functions $t_1$ and $t_2$:
$T=(t_1+t_2)/2$, $t=t_1-t_2$. In these equations, for shortness,
the arguments $({\bf k},T,t)$ of Green's functions ${\rm G}^{r,<}$
are dropped. $a_i$($i=1,2,3$) are the factors associated with the
directions of the momenta, $a_1=1+\cos (\chi_{\bf p}-\chi_{\bf
k})$, $a_2=1-\cos (\chi_{\bf p}-\chi_{\bf k})$, $a_3=\sin
(\chi_{\bf p}-\chi_{\bf k})$. We can see that in comparison with
the case of only the Rashba term\cite{Liu}, here, the factor $a_i$
is dependent on the momentum angle complicatedly through $\chi_{\bf
p}$. This fact leads to the kinetic equations analytically
unsolvable.

To further simplify our treatment, we adopt the generalized
Kadanoff-Baym ansatz (GKBA), which expresses the causality of the
time development of the two-time propagators using its equal-time
value\cite{GKBA,GKBA1}. At the same time, we neglect the collision
broadening of the retarded and advanced Green's functions and
retain the quantities only in the lowest order of gradient
expansion\cite{Jauho}. These simplifications correspond to the
Boltzmann limit and in result, our formalism becomes
quasiclassical. The kinetic equation for the distribution
functions can be followed
\begin{widetext}
\begin{equation}
\left [\frac{\partial}{\partial T}+e{\bf E}\cdot \nabla_{\bf
p}\right ]\rho({\bf p},T) -\frac {ie {\bf E}}{2}\cdot \nabla_{\bf
p} \chi_{\bf p}
[\rho,\sigma_x]+i\varepsilon_{RD}[\rho,\sigma_z]=-\left .\frac
{\partial \rho}{\partial T}\right |_{\rm scatt},\label{KE}
\end{equation}
where
\begin{equation}
\left .\frac {\partial \rho}{\partial T}\right |_{\rm
scatt}=\int_{-\infty}^T dt' \left [\Sigma^>{\rm G}^<+{\rm
G}^<\Sigma^> -\Sigma^<{\rm G}^>-{\rm G}^>\Sigma^<\right
](T,t')(t',T).
\end{equation}
\end{widetext}

Further, we assume that the dc field is sufficiently weak and only
the linear response in the steady state needs to be considered.
Otherwise, since we are just interested in the dissipationless
spin-Hall effect, it is sufficient to treat the nondiagonal
distribution functions in the lowest order of electron-impurity
scattering, {\it i.e.} in the order of $(n_i)^0$. Under these
considerations, the solutions of the kinetic equations (\ref{KE})
can be separated into two parts. The first part of solution
associates with the driving term $\frac i2e{\bf E}\cdot
\nabla_{\bf p}\chi_{\bf p}[\rho_0,\sigma_x]$ with $\rho_0$ being
the distribution function in equilibrium $\rho_0({\bf p})={\rm
diag}[n_{\rm F}(\varepsilon_1({\bf p})),n_{\rm
F}(\varepsilon_2({\bf p}))]$, and is a nondiagonal matrix in the
spin space. It can be written as
\begin{equation}
\rho_{12}^{(1)}({\bf p})=\rho_{21}^{(1)}({\bf p})=\frac{e{\bf E}\cdot {\nabla_{\bf p}\chi_{\bf p}}}{4\varepsilon_{RD}}\left \{n_{\rm F}[\varepsilon_1({\bf p})]-n_{\rm F}[\varepsilon_2({\bf p})]\right \},
\end{equation}
and results in the nonzero spin-Hall conductivity at zero
temperature\cite{Shen,Sinitsyn},
\begin{equation}
\sigma_{sH}^{(1)}\equiv \frac{{J_y^z}^{(1)}}{E}=
\left \{
\begin{array}{cc}
-e/8\pi\,\,\,&\alpha>\beta\\
0&\alpha=\beta\\
e/8\pi& \alpha<\beta
\end{array}
\right . .
\end{equation}

This part of spin-Hall conductivity corresponds to the intrinsic
spin-Hall effect and agrees with the previous studies in the
ballistic regime\cite{Shen, Sinitsyn}. Essentially, this
dissipationless spin-Hall effect stems from the dc-field-induced
transitions between two spin-orbit-coupled bands. When a dc field
is applied to the 2D system with spin-orbit coupling, electron
obtains an additional momentum, but its spin remains unchanged.
Consequently, in the helicity basis, the transition between the
spin-orbit-coupled bands occurs. It is obvious that all electrons
in 2D systems join in this process.

The second part of solution relates to the transport process and
the corresponding diagonal distribution functions should be
determined from the coupled equations
\begin{widetext}
\begin{equation}
-\left .\frac{\partial n_{\rm F}(E)}{\partial E}\right |_{E=\varepsilon_\mu} \frac{\partial \varepsilon_{\mu}}
{\partial p_x}=\pi \sum_{\bf k} |V({\bf p} -{\bf k})|^2 \{[\rho_{\mu\mu}^{(2)}({\bf p})-\rho_{\mu\mu}^{(2)}({\bf k})]
a_1\Delta_{\mu\mu}+[\rho_{\mu\mu}^{(2)}({\bf p})-\rho_{{\bar \mu}{\bar \mu}}^{(2)}({\bf k})]a_2\Delta_{\mu{\bar \mu}}\},
\end{equation}
with $\mu=1,2$, ${\bar \mu}=3-\mu$ and $\Delta_{\mu\nu}\equiv
\delta(\varepsilon_\mu({\bf p})-\varepsilon_\nu({\bf k}))$. And
then, the real part of nondiagonal distribution can be given by
\begin{equation}
{\rm Re}\rho_{12}^{(2)}={\rm Re}
\rho_{21}^{(2)}=\frac{\pi}{2}\sum_{{\bf k}\,\mu=1,2}|V({\bf
p}-{\bf k})|^2a_3(-1)^{\mu}
\{\Delta_{\mu\mu}[\rho_{\mu\mu}^{(2)}({\bf
p})-\rho_{\mu\mu}^{(2)}({\bf k})]-\Delta_{\mu{\bar \mu}}[\rho_{\mu
\mu}^{(2)}({\bf p})-\rho_{{\bar \mu}{\bar \mu}}^{(2)}({\bf k})]\}.
\end{equation}
\end{widetext}

This part of spin-Hall effect depends on the electron-impurity
scattering and entirely vanishes in clean samples. However, it is
of order of $(n_i)^0$ and the disorder only plays an intermediate
role. This disorder-mediated spin-Hall effect contributes from the
electrons near the Fermi surface and, physically, can be
understood as the following way. The electrons, participating in
the longitudinal transport, also can be scattered by impurities.
This process, which changes the electron momentum and conserves
its spin, yields the transition between two spin-orbit-coupled
bands in the helicity basis.

Unfortunately, the above coupled equations for diagonal
distribution functions only can be resolved numerically. We
perform a numerical evaluation for a GaAs-based heterojunction,
where the electron-impurity scattering is assumed to be
long-range\cite{Ando}. The momentum integration is calculated by
the Gauss-Legendre scheme. The linear system equations, arising
from the discretization of the integral equations for diagonal
distribution functions, are solved by using the singular value
decomposition. Here, we concern on the zero-temperature spin-Hall
effect. Therefore, only the distribution functions at the Fermi
surface need to be carried out.
\begin{figure}
\includegraphics [width=0.45\textwidth,clip] {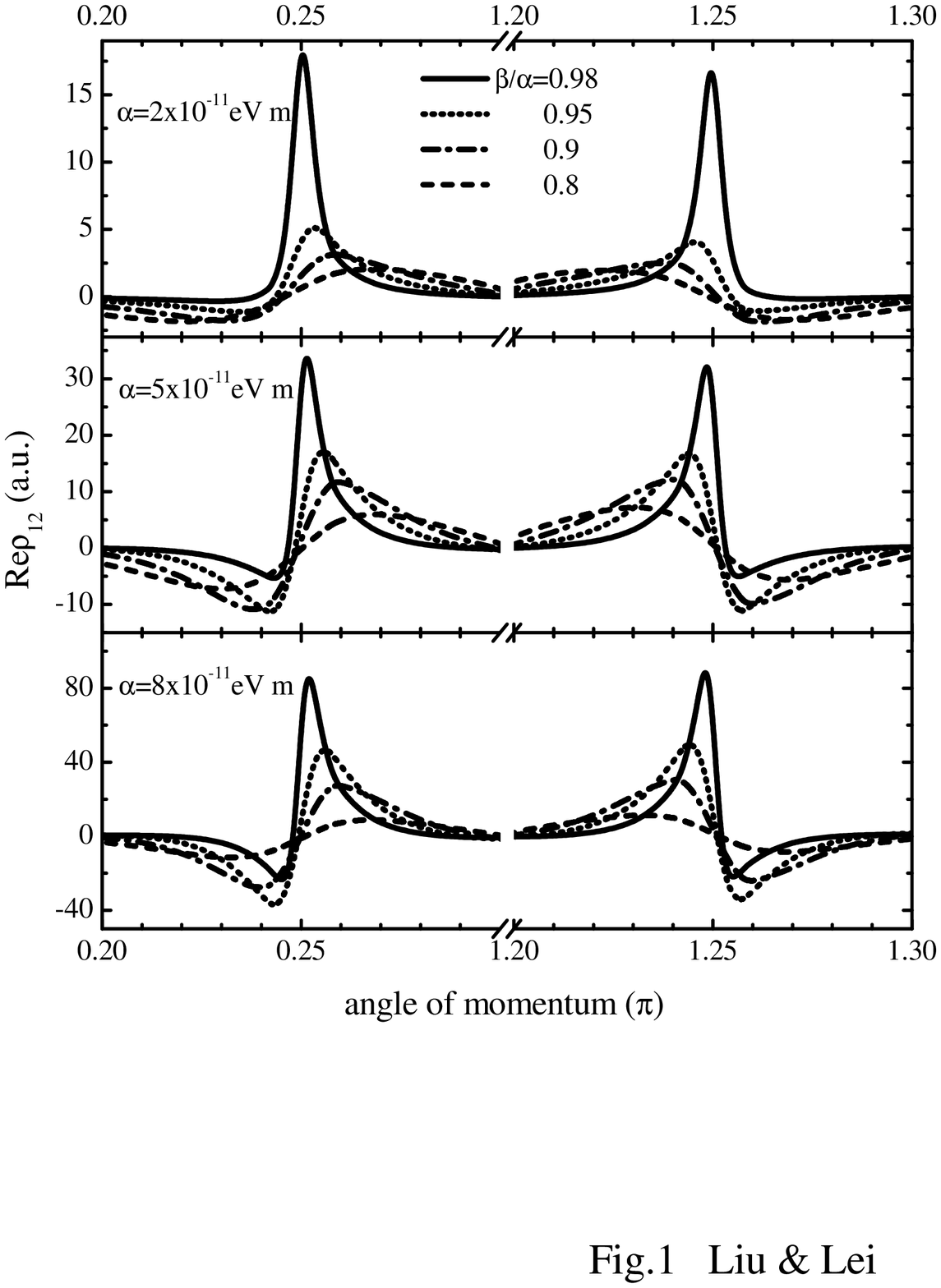}
\caption{The real part of total nondiagonal distribution functions $\rho_{12}$ near the angles $\pi/4$ and $5\pi/4$.
The electron density is $5\times 10^{11}$cm$^{-2}$.} \label{fig1}
\end{figure}

We know that, when only presence of the Rashba term, the
dependence of the nondiagonal distribution function on the
momentum angle can be expressed through the sine
function\cite{Liu}. However, in our case, the deviation from this
rule is expected due to the anisotropic dispersion relations. In
Fig.\,1, we plot the real part of nondiagonal distribution
functions, which could be observed by the experimental tools, such
as Raman spectroscopy, angle resolved photoelectron spectroscopy
(ARPES) {\it etc}. It can be seen that when the $\beta$ closes to
the value of $\alpha$, a peak always appears near the angle
$\pi/4$ or $5\pi/4$. With the ratio of $\beta/\alpha$ approaching
to the unit, the peak becomes more pronounced. The amplitude of ${\rm
Re} \rho_{12}$ for the same value of $\beta/\alpha$ increases with
the parameter $\alpha$. It is obvious that such structures near
$\pi/4$ and $5\pi/4$ relate to the zero of $\varepsilon_{RD}$ at
$\alpha=\beta$, producing a jump in the electron velocity.

Further, the spin-Hall current has been carried out by
substituting the ${\rm Re} \rho_{12}$ into Eq.\,(\ref{Jyz}). The
total spin-Hall current, coming from the above two parts of the
kinetic-equation solutions, is found to vanish. We have examined
this result in the wide range of parameters $\alpha$ and
$\beta\sim 1-10\times 10^{-11}$eV\,m. The typical electron
concentrations are taken to be $1-10\times 10^{11}$\,cm$^{-2}$. We
have seen, although the nondiagonal distributions are dependent on
the coupling constants and electron density, the total spin-Hall
effect in the estimated errors always becomes unobservable.

This complete cancellation of dissipationless spin-Hall effect
supports the phenomenological argument in Ref. \cite{Chalaev}: the
vanishing of spin-Hall current comes from the fact that the
uniform spin current is the total time derivative of the
magnetization. Here, when the presence of both the Rashba and
Dresselhaus spin-orbit couplings, the corresponding magnetization
operator can be taken as $\beta S_x+\alpha S_y$ with $S_i$ being
the spin operator. We should note that the nonexistence of
dissipationless spin-Hall effect is not a consequence of any
symmetries and is not necessarily suppressed to zero in
general\cite{Murakami}. It depends on the dispersion relation and
wave functions of the considered systems. For example, we have
carried out the spin-Hall conductivity for 2D hole systems and
found a nonzero dissipationless spin-Hall conductivity.

However, the dissipationless spin-Hall effect is expected to be
observed in the nonlinear regime. We can see from the kinetic
equations, the intrinsic part of spin-Hall conductivity always
linearly depends on the strength of dc fields. However, it is well
known that the transport of 2D electron systems becomes nonlinear
when the strength of dc field is of order of 1\,kV/cm. At that
time, the dc-field dependence of disorder-mediated part of
spin-Hall current, associating with the transport, is expected to
deviate from the linear rule. Hence, the non-dissipative spin-Hall
effect can be found for the strength of dc field larger than about
$1$\,kV/cm.

For 2D electron systems with both the Rashba and Dresselhaus
spin-orbit couplings, we even have calculated the longitudinal spin conductivity
$\sigma_{xx}^z$, defined in Ref.\,\cite{Sinova}.
Its value also has been found to be zero.

In summary, we construct a nonequilibrium Green's function
approach to the spin-Hall effect of 2D electron systems with both
the Rashba and Dresselhaus spin-orbit interaction. Taking into
account the long-range electron-impurity scattering, we have
resolved the kinetic equations numerically. It is found that,
although the dispersion relation of such systems exhibits an
anisotropic character, the total dissipationless spin-hall
conductivity vanishes.

One of the authors (SYL) gratefully acknowledge invaluable discussions with
Drs. W. S. Liu, Y. Chen and W. Xu. This work was supported by
the National Science Foundation of China, the Special Funds for
Major State Basic Research Project, and the Youth Scientific
Research Startup Foundation of SJTU.

\end{document}